# In Celebration of the Fixed Target Program with the Tevatron

June 2, 2000

Jeffrey A. Appel, Charles N. Brown, Peter S. Cooper, Herman B. White
Fermilab National Accelerator Laboratory, Batavia, Illinois

This document is an abridgement of the commemorative book prepared on the occasion of the symposium "In Celebration of the Fixed Target Program with the Tevatron" held at Fermilab on June 2, 2000. The full text with graphics contains, in addition to the material here, a section for each experiment including a "plain text" description, lists of all physics publications, lists of all degree recipients and a photo from the archives. The full text is available on the web at:

http://conferences.fnal.gov/tevft/book/

## *Table of Contents*

> *"The main application of the work here is spiritual, if you will. It's because, in a philosophical sense, in the tradition of Democritus, we feel we have to understand, in simplest terms, what matter is, in order to understand who we are."*

Robert R. Wilson, 1974

# 1. SYMPOSIUM PROGRAM

Symposium in Celebration of the
Fixed Target Program with the Tevatron
Friday, June 2, 2000

9:00 Session I - Convener: Jeffrey Appel

Getting to the Tevatron
Leon Lederman, Illinois Mathematics and Science Academy
and Illinois Institute of Technology

Theoretical Questions during the Tevatron Era
Jonathan Rosner, University of Chicago

10:10 Break

10:35 Session II - Convener: Chuck Brown

Evolution of Tevatron Accelerator and Experimental Technologies
John Peoples, Fermilab

Physics, Dectectors, and the Rest in the Tevatron Hyperon and Kaon Programs
Bruce Winstein, University of Chicago

11:45 Lunch

1:15 Session III - Convener: Peter Cooper

Physics of the Heavy Quark Program
Jussara Miranda, Centro Brasileiro de Pesquisas Físicas

Physics of the Muon and Neutrino Programs
Heidi Schellman, Northwestern University

2:25 Break

2:50 Session IV - Convener: Herman White

Studies of High – Pt and High – Mass Phenomena
Paul Slattery, University of Rochester

Summary of Where We Are and What Lies Ahead for Fixed Target at Fermilab
Michael Witherell, Fermilab

4:00 Celebration in the Wilson Hall Atrium

# 2. INTRODUCTION

The Tevatron is the world's first large superconducting accelerator. With its construction, we gained the dual opportunities to advance the state of the art in accelerator technology with the machine itself and the state of the art in particle physics with the experiments that became possible in a higher energy regime. In 1989 President Bush presented the National Medal of Technology to four Fermilab physicists; Helen Edwards, Dick Lundy, Rich Orr, and Alvin Tollestrup for their work in building the Tevatron. This award at the highest level possible for a government project is recognition of the many contributions from the Fermilab staff to the success of the Tevatron project; from Bob Wilson at the beginning through all the scientists, engineers, and staff of what we now call the Beams and Technical divisions. They turned Bob Wilson's vision into a real accelerator.

On the first of October 1983, the first run of the Tevatron, the still-largest superconducting accelerator in the world, was just beginning. The commissioning run, at 400 GeV/c with 5 fixed target experiments in the Proton and Meson Laboratories, was a stepping stone to higher energy for fixed target experiments and, a few years later, to the collider program.

At 00:40 the message on channel 13 read "NO BEAM TO PROTON FOR AT LEAST 3 HRS". In fact there were only 26 hours of beam delivered in that first month. This was quite frustrating to those sitting midnight shifts at the time, both at the experiments and in the accelerator control room.

However, it was also a tour-de-force in bringing together the technological frontier of industrial scale superconductivity and the energy, sensitivity and precision frontiers of particle physics. By mid January 1984, when this first run ended, the fixed target experimentalist's view of the accelerator was back to what it had always been - the black box from which the beam emerged. That attitude was a clear sign of success. The run was followed immediately by 800 GeV/c operations beginning April 1, 1984 with the same 5 beam lines running - two of them with new experiments.

There have been 43 experiments in the Tevatron fixed target program. Many of these are better described as experimental programs, each with a broad range of physics goals and results, and more than 100 collaborating physicists and engineers. The results of this program are three-fold: (1) new technologies in accelerators, beams and detectors which advanced the state of the art; (2) new experimental results published in the refereed physics journals; and (3) newly trained scientists who are both the next generation of particle physicists and an important part of the scientific, technical and educational backbone of the country as a whole.

Summaries of these results from the program as a whole are quite interesting, but the physics results from this program are too broad to summarize globally. The most important of the results appear in later sections of the book.

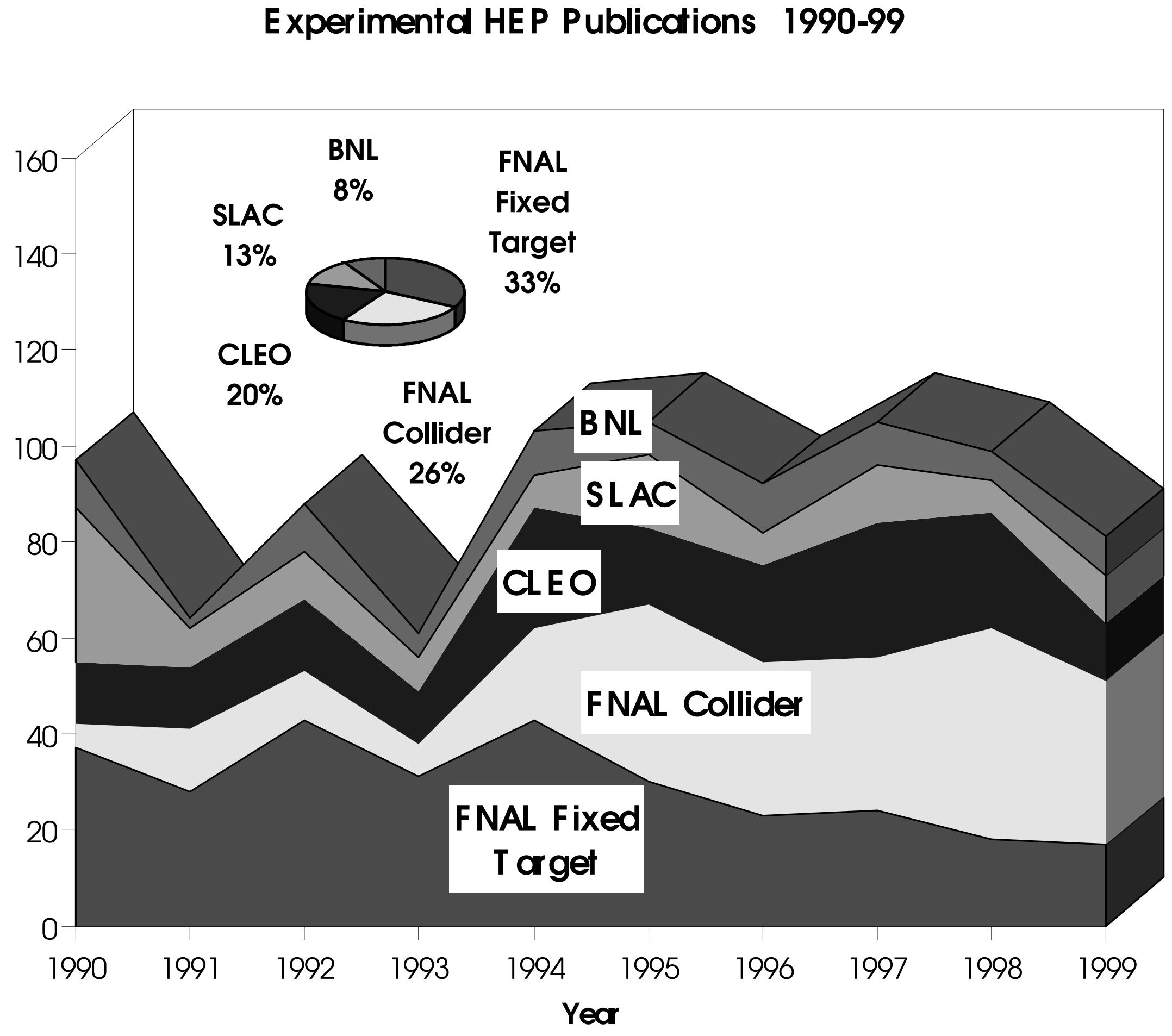

The papers and degrees lend themselves to a more quantitative programmatic analysis. We have counted the experimental papers resulting from all of the US laboratories over the decade of the 1990's which have been published in the major refereed journals. Of the 895 such papers, 294, one third, have come from the Tevatron fixed target program. We have counted a total 465 degrees, 381 doctorates and 84 masters degrees, earned by students from 104 universities based upon work done in the Tevatron fixed target program. For comparison, the Fermilab collider program, which started a few years later, has produced 310 advanced degrees. The international character of the fixed target program is evident when the advanced degrees are sorted by state and country. As the table below shows, universities in 30 of the US states and territories as well as 17 foreign countries gave advanced degrees for work done in this Fermilab program. Among the top 10 are Japan, Italy, Brazil and Germany. One third of all the advanced degrees were given by foreign universities. About one quarter of these degrees were earned by women.

| | MS | PhD | Tot | | MS | PhD | Tot |
|---|---|---|---|---|---|---|---|
| Belgium | - | 2 | 2 | Brussels University | - | 1 | 1 |
| | | | | Universiteit Antwerpen | - | 1 | 1 |
| Brazil | 8 | 13 | 21 | Centro Brasileiro de Pesquisas Fisicas | 5 | 8 | 13 |
| | | | | Federal University of Rio de Janeiro | - | 1 | 1 |
| | | | | Pontificia Universidade Catolica Rio de Janeiro | 1 | - | 1 |
| | | | | State University of Campinas | - | 1 | 1 |
| | | | | University of Sao Paulo | 2 | 3 | 5 |
| California | - | 26 | 26 | Stanford University | - | 1 | 1 |
| | | | | University of California at Berkeley | - | 3 | 3 |
| | | | | University of California at Davis | - | 7 | 7 |
| | | | | University of California at Los Angeles | - | 5 | 5 |
| | | | | University of California at San Diego | - | 1 | 1 |
| | | | | University of California at Santa Barbara | - | 7 | 7 |
| | | | | University of California at Santa Cruz | - | 2 | 2 |
| Canada | 2 | 7 | 9 | McGill University | 2 | 3 | 5 |
| | | | | University of Toronto | - | 4 | 4 |
| Colorado | - | 11 | 11 | University of Colorado | - | 11 | 11 |
| Connecticut | - | 7 | 7 | Yale University | - | 7 | 7 |
| England | - | 1 | 1 | Imperial College - London | - | 1 | 1 |
| Florida | - | 3 | 3 | Florida State University | - | 3 | 3 |
| France | - | 3 | 3 | University of Paris - Sud | - | 3 | 3 |
| Georgia | - | 1 | 1 | Georgia State University | - | 1 | 1 |
| Germany | 6 | 11 | 17 | Aachen University | - | 1 | 1 |
| | | | | Max-Planck-Institut Fur Kernphysik | 6 | 5 | 11 |
| | | | | Technischen Universitaat Munich | - | 2 | 2 |
| | | | | University of Freiberg | - | 2 | 2 |
| | | | | Wuppertal University | - | 1 | 1 |
| Greece | - | 5 | 5 | University of Athens | - | 5 | 5 |
| Hawaii | - | 1 | 1 | University of Hawaii | - | 1 | 1 |
| Illinois | 5 | 41 | 46 | Illinois Institute of Technology | - | 2 | 2 |
| | | | | Northern Illinois University | 5 | - | 5 |
| | | | | Northwestern University | - | 7 | 7 |
| | | | | University of Chicago | - | 19 | 19 |
| | | | | University of Illinois at Chicago Circle | - | 6 | 6 |
| | | | | University of Illinois at Urbana-Champaign | - | 7 | 7 |
| India | - | 3 | 3 | University of Delhi | - | 3 | 3 |
| Indiana | 3 | 16 | 19 | Ball State University | 2 | - | 1 |
| | | | | Indiana University | 1 | 5 | 6 |
| | | | | Notre Dame University | - | 11 | 11 |
| Iowa | 3 | 6 | 9 | University of Iowa | 3 | 6 | 9 |
| Israel | 2 | 2 | 4 | Tel Aviv University | 2 | 2 | 4 |
| Italy | 7 | 21 | 28 | University of Bari | - | 1 | 1 |
| | | | | University of Lecce | 2 | - | 2 |
| | | | | University of Milano | 4 | 9 | 13 |
| | | | | University of Pavia | 1 | 10 | 11 |
| | | | | University of Roma | - | 1 | 1 |

| | MS | PhD | Tot | | MS | PhD | Tot |
|---|---|---|---|---|---|---|---|
| Japan | 33 | 16 | 49 | Aichi University of Education | 1 | - | 1 |
| | | | | Kobe University | 3 | 2 | 5 |
| | | | | Kyoto University | - | 5 | 5 |
| | | | | Nagoya University | 4 | 2 | 6 |
| | | | | Osaka City University | 4 | 1 | 5 |
| | | | | Osaka University | 8 | 4 | 12 |
| | | | | Toho University | 5 | 2 | 7 |
| | | | | Utsunomiya University | 8 | - | 8 |
| Kansas | - | 3 | 3 | Kansas State University | - | 2 | 2 |
| | | | | University Kansas | - | 1 | 1 |
| Korea | 2 | 4 | 6 | Korea University | 2 | 4 | 6 |
| Maryland | - | 5 | 5 | University of Maryland | - | 5 | 5 |
| Massachusetts | - | 23 | 23 | Harvard University | - | 4 | 4 |
| | | | | Massachusetts Institute of Technology | - | 4 | 4 |
| | | | | Northeastern University | - | 8 | 8 |
| | | | | Tufts University | - | 5 | 5 |
| | | | | University Massachusetts at Amherst | - | 2 | 2 |
| Mexico | 4 | 6 | 10 | Cinvestav | 1 | 5 | 6 |
| | | | | Universidad Autonoma de San Luis Potosi | 2 | - | 2 |
| | | | | Universidad de Puebla | 1 | - | 1 |
| | | | | University of Guanajuato | - | 1 | 1 |
| Michigan | - | 16 | 16 | Michigan State University | - | 8 | 8 |
| | | | | University of Michigan | - | 8 | 8 |
| Minnesota | - | 7 | 7 | University of Minnesota | - | 7 | 7 |
| Mississippi | - | 1 | 1 | University of Mississippi | - | 1 | 1 |
| Missouri | - | 1 | 1 | University of Missouri | - | 1 | 1 |
| New Jersey | - | 10 | 10 | Princeton University | - | 4 | 4 |
| | | | | Rutgers University | - | 6 | 6 |
| New Mexico | - | 2 | 2 | New Mexico State University | - | 2 | 2 |
| New York | 1 | 32 | 33 | Columbia University | - | 16 | 16 |
| | | | | State University of New York at Albany | - | 1 | 1 |
| | | | | State University of New York at Stony Brook | - | 3 | 3 |
| | | | | University of Rochester | 1 | 12 | 13 |
| North Carolina | - | 5 | 5 | Duke University | - | 5 | 5 |
| Ohio | - | 11 | 11 | Case Western Reserve University | - | 1 | 1 |
| | | | | Ohio State University | - | 5 | 5 |
| | | | | University of Cincinnati | - | 5 | 5 |
| Oklahoma | - | 1 | 1 | University of Oklahoma | - | 1 | 1 |
| Pennsylvania | - | 15 | 15 | Carnegie Mellon University | - | 5 | 5 |
| | | | | Lehigh University | - | 1 | 1 |
| | | | | Pennsylvania State University | - | 2 | 2 |
| | | | | University of Pennsylvania | - | 3 | 3 |
| | | | | University of Pittsburgh | - | 4 | 4 |
| Puerto Rico | 6 | - | 6 | University of Puerto Rico | 6 | - | 6 |
| Russia | - | 1 | 1 | Moscow State University | - | 1 | 1 |
| South Carolina | - | 1 | 1 | University of South Carolina | - | 1 | 1 |
| Switzerland | - | 1 | 1 | University of Geneva | - | 1 | 1 |
| Taiwan | - | 1 | 1 | National Cheng-KunUniversity | - | 1 | 1 |
| Tennessee | - | 6 | 6 | University of Tennessee | - | 4 | 4 |
| | | | | Vanderbilt University | - | 2 | 2 |
| Texas | 2 | 13 | 15 | Rice University | 2 | 10 | 12 |
| | | | | Texas A&M University | - | 1 | 1 |
| | | | | University of Houston | - | 1 | 1 |
| | | | | University of Texas at Austin | - | 1 | 1 |
| Virginia | - | 5 | 5 | University of Virginia | - | 5 | 5 |
| Washington | - | 7 | 7 | University of Washington | - | 7 | 7 |
| Wisconsin | - | 8 | 8 | University of Wisconsin | - | 8 | 8 |

These numbers are incomplete. While data taking with 800 GeV/c beam ended in January 2000, analysis from some of the largest experiment programs will continue for several more years. There will probably be more than 50 additional papers published and a similar number of students graduated before this program will be truly complete.

Milestones denote progress on a journey. Now, seventeen years after the first beam was delivered, we have gathered to celebrate the milestone of the end of fixed target runs with the Tevatron. The symposium will look at where we have been and where we will be going next. The physics results from the Tevatron fixed target program have helped form our present understanding of nature and shaped some of the directions for future research. The technological advances we have made have already found their way into our new Main Injector and other accelerators, as well as current and planned future generations of detectors. The scientists who were trained in the Tevatron fixed target program, either as graduate students or in more senior positions, form the core of the upcoming Main Injector fixed target program. That program is already digging holes in the ground for future neutrino experiments, new experiments but with the same types of beams that began our fixed target programs in the past.

The symposium is held not to bury fixed target physics or simply to celebrate it, but to remind ourselves of the exceptional progress we have made, and to begin again with a next generation of Fermilab fixed target experiments. As with Mark Anthony, we'll sneak in a little praise for the old program as well.

## 3. INVESTIGATING QCD AT FIXED TARGET ENERGIES

E609 - The Structure of High $P_T$ Hadronic Interactions

E683 - Photoproduction of High $P_t$ Jets

E690 - Study of Charm and Bottom Production

E704 - Experiments with the Polarized Beam Facility

E705 - Charmonium and Direct $\gamma$ Production at 300 GeV/c

E706 - Direct $\gamma$ Production in Hadron Induced Collisions

E711 - Dihadron Production

E772 - The Quark-Antiquark Sea in Nuclei

Quantum Chromodynamics (QCD) describes the strong force that binds quarks within hadrons, the class of strongly interacting particles that includes the proton and the neutron. QCD has some similarities to QED (Quantum Electrodynamics), the quantum theory of the electromagnetic force. However, there are substantial differences that affect our ability to make calculations using these theories. Calculations based on QED are the most accurate in all of physics, while those involving QCD can often only be approximated. These differences arise partly from the charge structure of the two theories. QED has a single charge, the familiar electric charge. QCD has three types of charge, referred to as "colors" (the "Chromo" in Chromodynamics), which combine to yield neutral (colorless) states in analogy to the way three primary colors combine to create white, or colorless, light.

As a consequence of the existence of three types of QCD charge, instead of being mediated by a single massless particle (the photon of QED), QCD involves eight massless intermediaries, known as gluons (that "glue" quarks together). Moreover, while a photon is electrically neutral and does not interact directly with other photons, gluons carry color charge and interact strongly with one another.

The QCD force is about 100 times stronger than QED, and includes more complicated interactions than QED (because of the gluon-gluon interaction). In addition, although the QED force between charged particles becomes weaker with distance, at typical intra-quark distances in hadrons the QCD force between quarks is approximately constant. Thus the energy in the color field increases linearly with the distance between the quarks (until there is enough energy in the field to create quark-antiquark pairs). This property leads to the confinement of quarks within ordinary (colorless) matter, enabling the weaker QED force to dominate on atomic scales (with the happy results that atoms have the structure they do and that we exist to comprehend such things as QED and QCD).

Quark confinement also makes QCD more difficult to study at fixed target energies. This follows from the uncertainty principle, distance and momentum measurements are inversely related (large distances result in small amounts of momentum being transferred between interacting particles, at small distances large amounts of momentum can be exchanged). Thus, a mathematical technique known as perturbation theory, which is most reliable under circumstances in which an interaction is weak, can be employed to carry out very precise QED calculations in the kinematic regime of large distances and low momentum transfers. These conditions are relatively easy to access at fixed target energies. In contrast, the QCD force is weak only at small distances and large momentum transfers, a regime most easily reached in the highly energetic collisions characteristic of collider interactions.

Studying QCD at fixed target energies must necessarily involve experiments that address simple interactions that can be calculated reasonably accurately at lower energies, or that focus on processes that benefit from the increased precision of very high event rates achieved in fixed target experiments.

# 4. STRUCTURE OF PROTONS, NEUTRONS, AND MESONS

E605 - Leptons and Hadrons Near the Kinematic Limits

E615 - forward production of muon pairs

E665 - Muon Scattering with Hadron Detection

E733 - The Study of High Energy Neutrino Interactions with the Tevatron Quadrupole Triplet Beam

E744/770 - Neutrino Physics at the Tevatron

E745/782 - Neutrino Experiment using the one-meter High-resolution Bubble Chamber

E866 - Measurement of anti-D(x)/anti-U(x) in the Proton

There are some ways in which finding out what's inside a proton or neutron is analogous to finding out what's inside the nucleus of an atom: fire probing particles (say, electrons) of known energy at the nucleus, and look at the energy and angle of the scattered particles - a standard sort of "fluoroscopy by bombardment" technique. The electromagnetic interaction between the charge of the probe and the charge distribution of the nucleus will reveal how the target charges are arranged. If the probe energy is high enough, the target will disintegrate, and the fragments, their energies, and their angles, can be analyzed for clues to the composition and structure of the target.

In many ways, however, the story isn't quite so simple when it comes to nucleons and mesons. As the distance scale being probed decreases (with increasing probe energy), the fluoroscopy "evolves" in a way different from that of the atomic nucleus.

When a nucleon is probed at the greatest distance scale, it responds as a single object. At shorter lengths (higher probe energies), the momentum distribution of the three valence quarks is revealed and some of the momentum of the nucleon is observed to reside in the quanta of the color force, the gluons. At the smallest distance scales (highest probe energies), still more momentum carriers manifest themselves – a sea of quark-antiquark pairs coupled directly to the gluons. The momentum distributions of these constituent particles (valence quarks, sea quarks and antiquarks, and gluons, collectively called partons) are called the parton distribution functions. They are derived from the interaction probability (the cross-section) of the probe lepton as a function of energy and angle.

The parton distribution functions parameterize the input hadron structure distributions in the strong interaction description of lepton-hadron and hadron-hadron interactions. As such, they may also be investigated by studying the dynamics of many processes; e.g., single hard-photon production at high transverse momentum, the production of high-mass pairs of muons, and the production of heavy quarks. To the extent that these other processes are described by the same distribution functions, QCD is providing a reliable description of strong interactions. These other processes can then also be incorporated in the global fits to produce a better determined set of distribution functions. While most of the relevant measurements appear in experiments in this chapter, others can be found in Sections 3 and 5.

The universality of the parton distribution functions provides evidence for the reality of partons, objects which have never been observed in isolation (QCD predicts that this can never happen). The parton distribution functions form the backbone of our understanding of partons, and provide the basis for the predictive power of the Standard Model (and the base for searches for new phenomena that lie outside the Standard Model).

# 5. PHYSICS OF CHARM AND BEAUTY

E400 - charmed particle production by neutrons

E653 - Charm and Beauty Decays in a Hybrid Emulsion Spectrometer

E672 - Hadronic Final States in Association with High Mass Dimuons

E687 - Photoproduction of Charm and Beauty

E691 - Charm Production with the Tagged Photon Spectrometer

E743 - Charm Production in pp Collisions with LEBC-FMPS

E769 - Hadroproduction of charm

E771 - Beauty Production by Protons

E781/SELEX - Study of Charm Baryon Physics

E789 - beauty-Quark Mesons and Baryons

E791 - Hadroproduction of Charm

E831/FOCUS - Heavy Quarks study Using the Wideband Photon Beam

Two of the fundamental particles of the standard model are the charm quark and the beauty quark, the latter sometimes known as the bottom quark. In spite of their quaint names, these quarks have both played, and continue to play critical roles in particle physics research. The charm quark was the real spark for the acceptance of the whole picture of the quark-lepton sub-structure of matter. The bottom quark is now the focus of extensive studies around the world, studies which aim at understanding the details of standard-model CP violation and of the matter-antimatter asymmetry observed in experiments. More hopefully, these studies may find a glimpse of what lies beyond the standard model. Fermilab fixed-target experiments on particles containing charm quarks have provided guideposts in the understanding of CP violation. Given that the standard model explanation for the CP violation seen in the laboratory cannot explain the mysterious asymmetry in the matter-antimatter balance in the universe, there must be something beyond the standard model. Charm and beauty studies may hold the key to this and other tantalizing questions in particle physics. The mixing of charm particle and antiparticle, and the searches for rare and forbidden decays may open unexpected doors. Also, we may hope that a detailed study of physics involving charm and bottom quarks, when combined with the study of the other quarks, will lead to an understanding of why nature has arranged the quarks into three generations, with each generation containing two sometimes quite different quarks.

Even in the area of standard-model physics, there are important questions to which fixed-target experiments at Fermilab have contributed answers: how quarks are produced in high-energy interactions, how those quarks turn into the particles seen in the laboratory, and the dynamics leading to the decay of particles containing charm quarks. For example, since the charm and bottom quarks are produced dominantly by the fusion of bits of the glue that binds quarks within the particles that are seen directly in the laboratory, charm quark production properties may be used to study the distribution of the glue in particles. Also, the decay of charm particles provides a particularly clean environment in which to study the characteristics of those particles into which the charm particles decay. The recent charm experiments are providing answers that have eluded physicists for decades. The copious decays observed in Fermilab fixed-target experiments provide a unique way of studying the low-mass resonances of pairs of pions and of pions and kaons. These measurements complement those made historically in scattering experiments at lower energies.

Major contributions from the charm and beauty fixed target program also have been in the areas of detector development, and data acquisition and computing. The rarity of charm and beauty quarks in fixed target interactions and the unique decay properties of these quarks led experimenters to implement silicon microstrip detectors, trigger processors, fiber readout of scintillating plastics, high speed data readout, and web-based monitoring. Creativity, trying non-standard techniques, and diverse and innovative beams have marked this fixed target program. Each experiment tried a different wrinkle to advance the science.

# 6. SYMMETRY TESTS

E621 - Measurement of the CP Violation Parameter $\eta_{+-0}$

E731 - A Precision Measurement of the CP Violation Parameter ($\varepsilon'/\varepsilon$) in the Neutral Kaon System

E773 - Measurement of the Phase Difference Between $\eta_{00}$ and $\eta_{+-}$ to a Precision of 1°

E774 - Electron Beam Dump Particle Search

E799 - Rare Decays of $K_L^0$ and Hyperons

E832/KteV - A Search for Direct CP Violation in $K_L^0 \rightarrow 2\pi$

E871 - Search for CP Violation in the Decays of $\Xi^- / \bar{\Xi}^+$ and $\Lambda / \bar{\Lambda}$ Hyperons

Objects in our everyday macroscopic world are often said to possess symmetry. We say this when there are operations on an object that could alter its appearance, but in fact do not; an example of this is the rotation of a sphere through an angle about its center. Symmetry concepts can be applied not only to the physical processes that govern our local, large scale world, including such things as plant growth, the composition of rocks, and the flight of space ships, but also to the microscopic world.

To fully understand nature`s principles, it is believed that studying the relationship between symmetry and the physical laws is best done at the most fundamental level, the level of quarks and leptons. Over the last few decades, the field of elementary particle physics has made tremendous progress in understanding the nature of fundamental interactions and fundamental structure. New discoveries have propelled us toward a substantial strengthening of the Standard Model of particle physics, while advanced technologies have enhanced our ability to investigate the compelling questions of the origins of mass, matter and antimatter in the universe, and the basic nature of symmetry. Symmetry arguments have a long and extensive history, including the history of many of the basic principles of physical laws. And, with the introduction of successful quantum field theories and more mathematically rigorous techniques over the decades, symmetry principles have been used as tools to help define and extend natural laws. Also, experimental data has been studied for the accuracy of compliance with symmetry principles; and when symmetry is violated, to what degree and by what mechanisms.

The experiments that operated during the Tevatron fixed target era extended the investigation of the relatively old symmetry in nature known as CP, including the search and study of extremely rare decay processes. The C in this mathematical formulation represents the independent symmetry that describes the interchange of matter and antimatter. The P in this combined operation represents the independent symmetry of parity, an operation that changes left to right (like looking in a mirror). The combined operation, CP, was thought to be invariant in particle interactions, i.e., the interactions would be symmetrical in CP terms. However, in some very rare cases, this symmetry is violated. The masses and interactions of particles are nearly identical to those of their corresponding antiparticles, but there is a small difference in this CP symmetry of nature, only observed so far in the decays of K mesons. Since matter and antimatter mutually annihilate, and (at creation) our universe is postulated to include equal amounts of matter and antimatter, this very small CP difference may contribute a crucial bit of information that helps explain the abundance of matter over antimatter in the universe today. Although an ongoing program, differences in the accuracy of the experimental measurements of these processes world wide has provided another motivation for new studies. A number of experiments presented here had a goal, not of just observing this well established symmetry violation, but to measure the parameters of these very rare processes to high precision. The results could improve our understanding of these phenomena in the context of the Standard Model, which so successfully describes many aspects of elementary particle interactions.

These experiments also provided some of the most versatile detector instruments built during this era, making it possible also to study hyperon physics, to make particle lifetime measurements, and to search for supersymmetric particles. Forethought in the design of the beams, particularly in the kaon experiments, also provided for the creation of exceptional, secondary, neutral hyperon beams that allowed dedicated studies of hyperons produced at high energies. The scope of these studies also includes the search for violation of well established symmetries such as CPT, where one includes the operation of time reversal invariance T and searches for rare, short lived new particles. The results gained from the successful study of CP violation in the decay of the neutral kaon, does not answer one intriguing question; why is this symmetry violation seen only in the kaon system? Many non-kaon experiments around the world are currently underway to search for this symmetry violation in other systems. Experiments are underway on CP violation in hyperon decay at Fermilab and on B meson decay here and at other laboratories – with high priority.

The Fermilab fixed target symmetry experiments produced a number of high precision, rare decay measurements, including first measurements of some decay modes. One such measurement observed the first CP violating effect in an angular distribution variable, and included demonstrated CP- and T-odd effects in the asymmetry. From the successful early measurements of CP violating symmetries to unprecedented precision, to the latest fixed target experiments and the results of further analysis to follow, these experiments are well on their way to completing their goals.

# 7. HYPERONS AND NEUTRINOS

E632 - An Exposure of the 15' Bubble Chamber with a Neon- Mixture to a Wideband Neutrino Beam from the Tevatron

E715 - Precision Measurement of $\Sigma^- \rightarrow n e^- \nu$

E756 - Magnetic Moment of the $\Omega^-$ Hyperon

E761 - An Electroweak Enigma: Hyperon Radiative Decays

E800 - High Precision Measurement of the $\Omega^-$ Minus Magnetic Moment

E815/NuTev - Precision Neutrino / Antineutrino Deep Inelastic Scattering Experiment

E872 - Measurement of $\tau$ Production from the Process $\nu_\tau + N \rightarrow \tau$

In the beginning, at Fermilab, there were neutrinos and hyperons created by beams from the Main Ring. Among the first experiments proposed, approved, and run were E1A, and E21, the first generation of Fermilab electronic neutrino detectors, and E8 the first generation Fermilab hyperon beam experiment. There were also several bubble chamber exposures in the neutrino beam. These approaches matured into complete programs of experiments in the Tevatron fixed target era.

The common physics thread shared by these two rather different beam particles is the weak interaction. Neutrinos are particles which feel only the weak force. This makes them excellent probes of complicated structures, like protons and neutrons. Some of the Tevatron neutrino experiments were so sharply focused in the area, that they have been included in the section on proton, neutron, and meson structure rather than here. Neutrinos are also excellent places to study the weak interaction itself. They don' t do anything else, so they are a clean weak interaction laboratory. Neutrino scattering cross-sections (the probability that they will actually hit something in a given target) increase linearly with the neutrino energy. This made the 400 GeV Fermilab Main Ring good for doing neutrino physics, and the 800 GeV/c Tevraton even better.

The hyperons are the particles in the same family as the proton and the neutron, but containing one or more strange valence quarks. Only the weak interaction does not conserve strangeness; it is the only way a hyperon can decay. This makes hyperons live $10^{14}$ times longer than their non-strange cousins, the excited non-strange proton and neutron states. This is long enough to make hyperon beams that will go many meters at Tevatron energies before most of the hyperons decay. Hyperons, like protons, are particles of spin ½. This makes it possible to have polarized hyperon beams; something which is impossible with a spin 0 K meson beam. Polarization is a delicate and sensitive probe of both the weak interaction controlling the hyperon`s decay and the structure of the quarks and other stuff which makes up the hyperon itself. E8 discovered in 1976 that hyperons were produced with significant polarization. Tom Devlin, of Rutgers University, and Lee Pondrom, of the University of Wisconsin, were subsequently awarded the Panofsky prize for this discovery and the sequence of experiments it enabled.

Typically only one process happens at a time in weak interactions, one quark decays to another, or a neutrino hits just one quark in a target proton. The combination of the cleanliness of the weak interaction and the high intensity beams of both neutrinos and hyperons available at the Tevatron allowed a set of experiments of unprecedented precision.

A carefully crafted experiment can isolate just one particular aspect of the structure of a proton in order to study it carefully. For example, the E632 dimuon result focused in on the charmed quark content of the proton - which only exists in the *sea* of virtual quark antiquark pairs inside the proton. In a similar but different example, in a series of experiments all the hyperon magnetic moments were measured with high precision, including the $\Omega^-$ (by E800) which is made of three strange valence quarks. The results are sufficiently precise to both confirm our basic understanding of the structure of the baryons (the family to which protons, neutrons and hyperons belong) and to confound theoretical description anywhere close to the present experimental uncertainties.

E815/NuTeV will make a precision measurement of the Weinberg angle, $\sin^2(\theta_W)$, a fundamental electro-weak parameter more normally associated with the very high energy scales of $e^+e^-$ and hadron colliders. E872 is seeking to observe the last fundamental fermion, the $\tau$ neutrino.

# 8. TECHNOLOGICAL DEVELOPMENTS

## 8.1 INTRODUCTION

The physics produced by the Tevatron fixed target program depended on a whole sequence of technological developments in accelerators, detectors, and computing. Many experiments required one or more of these innovations to even be feasible in the first place, and to achieve their goals. The experimental collaborations themselves contributed significant advances in the technologies required. These advances have had far-reaching impact on subsequent work, both at Fermilab and elsewhere in the world. Some of these advances transcend in importance the experiments for which they were developed. In this section we will hit a few of the highlights in this area.

## 8.2 ACCELERATOR AND BEAM LINES

By far the most impressive, important, and obvious technical development is the Tevatron itself. From the experimentalist`s viewpoint, it not only provided twice the beam energy but also a much longer spill. The spill duty factor was increased from 1 second of beam every 15 second to 20 seconds of beam every 60 seconds; a factor of five improvement over the Main Ring. The longer spill length required several important innovations in the Tevatron itself. An extraction system, generically known as QXR, was extensively modified based upon microprocessor technology to handle the much longer beam spill.

The external beamlines required significant new developments as well. The longer spill implied much less intensity per second in the external beamlines. A new beam position monitoring system was developed to sense and control the external beams. These beams are as much as one million times less intense than the circulating beam in the Tevatron itself. The higher energy beam required twice the bending power in existing tunnels, which were originally designed for 200 GeV beams. The new requirement was to deliver the extracted 800 GeV protons to the Meson, Proton, and Muon Laboratories. This led to the development of the strings of superconducting magnets and their associated cryogenic plants, known as the left, right and muon bends. Many will remember channel 13 messages like " LEFT BEND QUENCH – NO ESTIMATE" - these things were far from trivial.

The higher beam energy and the sensitivity of the downstream cryogenic magnet strings required significant upgrades to the stations that split the extracted proton beams to each of the Proton, Neutrino, Muon, and Meson Laboratories. Motion controls and new pulsed magnets were installed to move the fast spill around the septa without burning them up, and the septa themselves were improved.

## 8.3 DETECTORS

The most notable and noted advance in detector technology was the development of the silicon microvertex detectors by the Santa Barbara group for the charm photoproduction experiment E691. This development, built on the detector development work of European groups working at CERN, revolutionized heavy quark physics. This occurred first in charm production, and latter in studies of hadrons with beauty in both fixed target and collider experiments, as well as in the discovery of the top quark in the Tevatron collider . This development received significant recognition with the 1990 award of the Panofsky Prize to Mike Witherell for the advances in charm physics that it made possible.

The group from the Petersburg (then Leningrad) Nuclear Physics Institute made a major advance in precision electron identification with their development of a large transition radiation dectector (TRD) system for the precision $\Sigma^-$ beta decay experiment E715. This large system achieved electron/pion separation of several thousand, with an electron inefficiency of less than 1%. These techniques have been used and further extended at Fermilab in experiments E761, E781, and E799.

The KTeV experiment developed a CsI photon calorimeter with outstanding energy and position resolution, excellent linearity, and very high rate capabilities. They achieved better than 0.75 % energy resolution over virtually their entire photon energy range of 5-100 GeV. They developed a new digital readout system with 17 bits of dynamic range housed in the photomultiplier base of each of their 3100 CsI crystals.

The heart of The KTeV CsI readout was an application specific integrated circuit (ASIC) called the QIE chip, one of several ASIC' s developed on the 14$^{th}$ floor of Wilson Hall for use in the Tevatron fixed target program. Others included silicon microstrip and wire chamber electronic circuits. These chips, and the ability to develop new circuits in the latest technologies, have come into widespread use since these early efforts.

The technique of recording the position of Cerenkov photons at the focal plane of a large Cerenkov detector for particle identification over a broad angular range (the Ring Imaging Cerenkov Counter, RICH) was pioneered by the E605 experiment. A Fermilab group lead the effort in SELEX (E781) to develop a large ring imaging Cerenkov counter (RICH) based upon 2848 small phototubes as the photodetector. The SELEX RICH achieved useful $\pi$-K separation in full multi-hadronic events up to 165 GeV/c with 12 photons observed on a typical ring. The new CKM experiment planned for the Main Injector fixed target program is based, in part, on this detector technique that accommodates very high beam rates with excellent time resolution.

## 8.4 COMPUTING

The use of `farms` of parallel computers bases upon commercially available processors is largely an invention of the Fermilab Advanced Compter Project (ACP). This technique has become widespread, both for on- and off-line computing - for selecting, acquiring, and processing the vast volumes of data that come with a full hardronic cross-section.

The original ACP I computer was developed at Fermilab in the mid 1980`s based on a Motorla 68020 processor and an ACP designed bus structure which allowed dozens of processors to analyze individual events in parallel. Much of the data from the early Tevatron fixed target runs were reconstructed off-line on farms of ACP I processors in the Feynman Computing Center. At its peak, we had 400 processors in use.

This technique was extended in the next generation to farms to the use of commercial UNIX workstations. Farms like these have been established in the Feynman Computing Center as well as collaborating universities and national laboratories. For example the 50 terabytes of data colleted by experiment E791 was reconstructed in parallel on farms at Fermilab, the University of Mississippi, Kansas State University and the CPBF in Rio de Janeiro.

This is an innovation which has become an industry standard in our field. Now there is hardly an experiment that does not have both an on-line computing farm for sophisticated software trigger decisions and an off-line farm for rapid parallel reconstruction of events.

As the Tagged Photon Laboratory charm group moved from experiment E769 to experiment E791, they pioneered the use of 8 mm magnetic tape for online data recording. The change to a new recording medium was adopted as a new standard by the Computing Division (then the Computing Department in the Research Division). After considerable work (and pain), it became the de facto standard for data recording, both at Fermilab and in many other places in particle physics.

# 9. ACKNOWLEDGMENTS

The book would not have been possible without the help of a large number of people. The many spokespeople from the experiments in the full book, and their designees, have provided us with most of the material presented in the full book. We also want to express our sincere appreciation to Sue Schultz for her lead in the typing, compiling, and formatting the book, as well as for putting up with all the variations of input and the diverse requests from the editors. We appreciate the encouragement of Judy Jackson who pushed for the "plain English" texts which have been prepared for each section (reproduced here) and experiment, and the help of Jackie Coleman in the Directorate.

We thank Paul Nienaber and Paul Slattery who contributed material to the summaries. Finally, we thank those who contributed material to the full book. They include: R. Brock, D. Christian, M. Corcoran, B. Cox, M. Crisler, J. Cumalat, C. Dukes, G. Ginther, G. Gollin, N. Grossman, J. Hanlon, Y-B Hsiung, K. Johns, K-B Luk, B. Lundberg, K. McDonald, M. Purohit, S. Reucroft, H. Schellman, W. Selove, M. Shaevitz, W. Smart, N. Stanton, B. Winstein, Y. Wah, J. Wiss, A. Yokosawa, and A. Zieminski.

# APPENDIX - PROGRAM HISTORY

**HISTORY OF TEVATRON OPERATION**

**1983 - 1988**

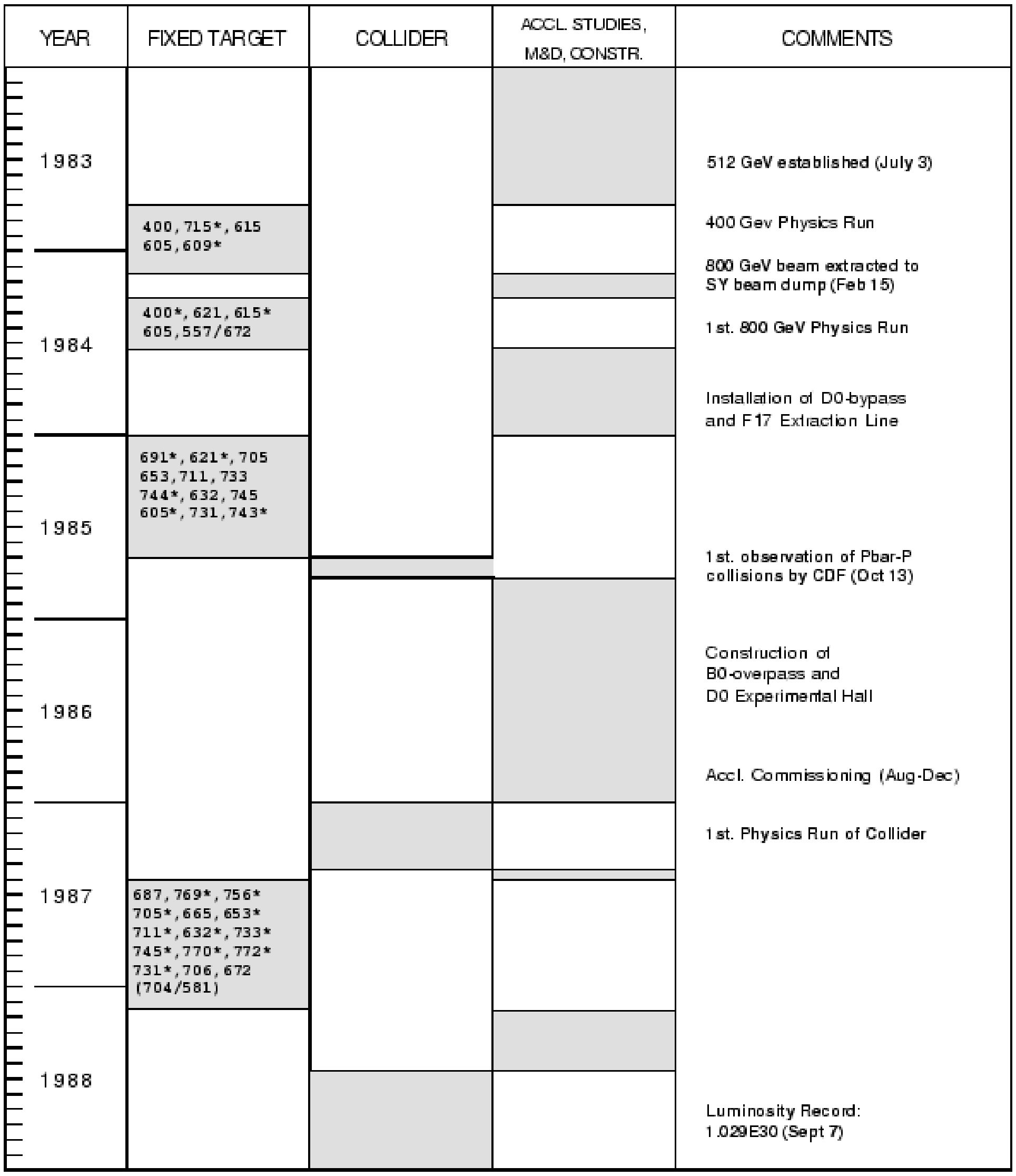

## HISTORY OF TEVATRON OPERATION

## 1989 - 1994

| YEAR | FIXED TARGET | COLLIDER (Accumulator) | ACCL. STUDIES, M&D, CONSTR. | COMMENTS |
|---|---|---|---|---|
| 1989 | | 710*, 713*<br>735*, 741*<br>778 | | 778 run only two weeks<br>at the end |
| 1990 | 687, 774*, 683<br>791, 761*, 771<br>665, 690, 782*<br>789, 704*, 773<br>706, 672 | (760) | | Mid February to<br>End of August<br>6.5 months |
| 1991 | | 778* | | A short run for SSC related<br>accelerator study |
| | 683*, 687*, 791*<br>800*, 771*, 665*<br>690*, 789*, 773*<br>799, 672*, 706* | (760*) | | Mid July to Mid January<br>6 months |
| 1992 | | | | Accelerator Startup<br>(Mid May through August) |
| | | 740, 775 | | September 92 to May 93<br>9 months |
| 1993 | | | | Accelerator Startup |
| 1994 | | 740, 775, 853 | | 12/15/93 to 8/26/94<br>8.5 months |
| | | 740, 775, 811<br>853, 868 | | |

* Completed

# HISTORY OF TEVATRON OPERATION

## 1995 - 2000

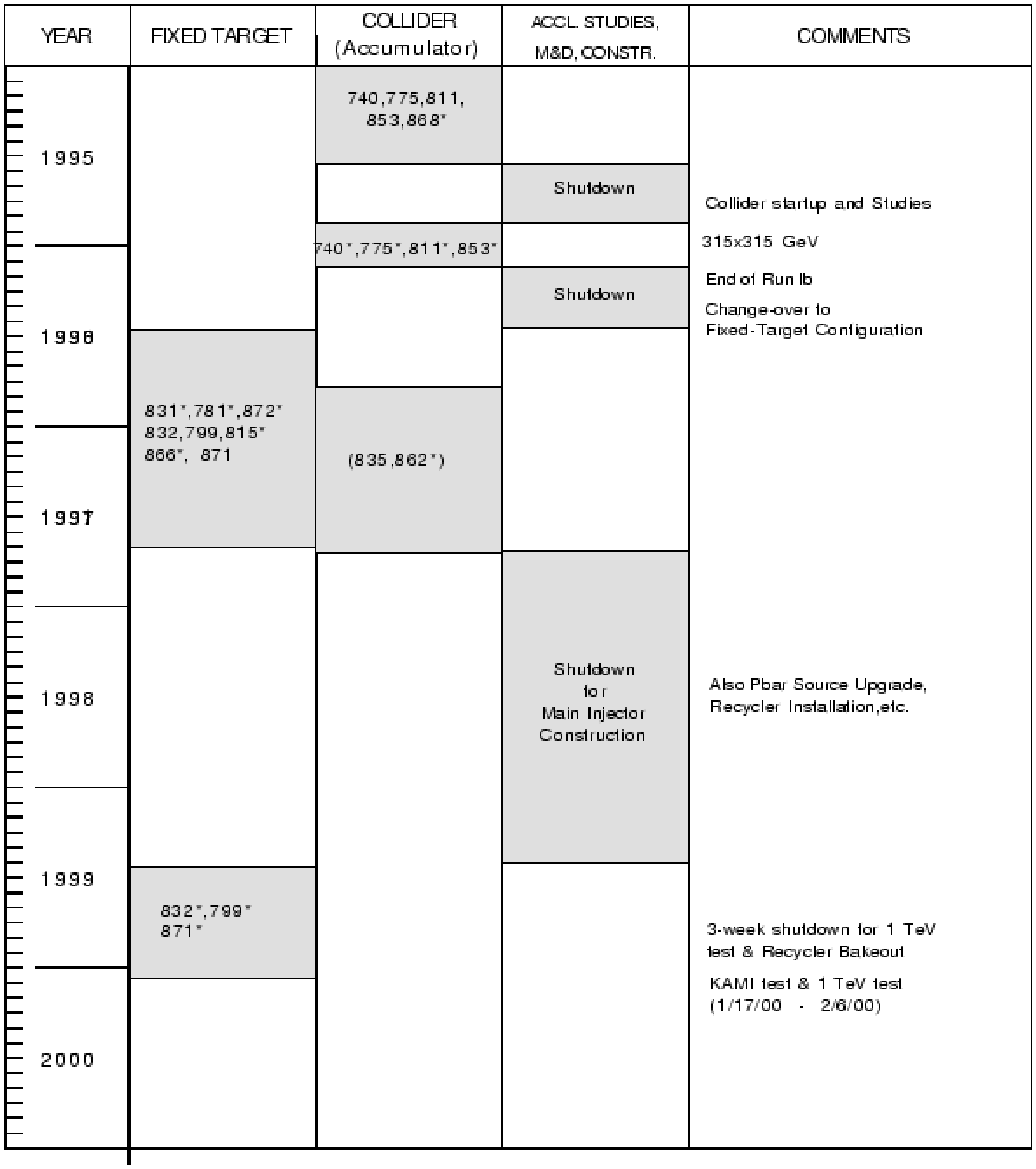

* Completed